\documentclass[english,aps,manuscript,aps,manuscript,showpacs]{revtex4}
\usepackage[T1]{fontenc}
\usepackage[latin9]{inputenc}
\usepackage{float}
\usepackage{amsmath}
\usepackage{amssymb}
\usepackage{graphicx}

\makeatletter

\newcommand{\noun}[1]{\textsc{#1}}

\@ifundefined{textcolor}{}
{%
 \definecolor{BLACK}{gray}{0}
 \definecolor{WHITE}{gray}{1}
 \definecolor{RED}{rgb}{1,0,0}
 \definecolor{GREEN}{rgb}{0,1,0}
 \definecolor{BLUE}{rgb}{0,0,1}
 \definecolor{CYAN}{cmyk}{1,0,0,0}
 \definecolor{MAGENTA}{cmyk}{0,1,0,0}
 \definecolor{YELLOW}{cmyk}{0,0,1,0}
}

\makeatother

\usepackage{babel}
\begin{document}

\title{Acceleration of spin-orbit coupled Bose-Einstein condensates: analytical
description of the emergence of Landau-Zener transitions}

\author{J.M. Gomez Llorente and J. Plata}

\address{Departamento de F\'{\i}sica, Universidad de La Laguna,\\
 La Laguna E38204, Tenerife, Spain.}
\begin{abstract}
We analytically study the effect of gravitational and harmonic forces
on ultra-cold atoms with synthetic spin-orbit coupling (SOC). In particular,
we focus on the recently observed transitions between internal states
induced by acceleration of the external modes. Our description corresponds
to a generalized version of the Landau-Zener (LZ) model: the dimensionality
is enlarged to combine the quantum treatment of the external variables
with the internal-state characterization; additionally, atomic-interaction
effects are considered. The emergence of the basic model is analytically
traced. Namely, by using a sequence of unitary transformations and
a subsequent reduction to the spin space, the SOC Hamiltonian, with
the gravitational potential incorporated, is exactly converted into
the primary LZ scenario. Moreover, the transitions induced by harmonic
acceleration are approximately cast into the framework of the basic
LZ model through a complete analytical procedure. We evaluate how
the validity of our picture depends on the system preparation and
on the magnitude of atomic-interaction effects. The identification
of the regime of applicability and the rigorous characterization of
the parameters of the effective model provide elements to control
the transitions.
\end{abstract}


\maketitle

\section{Introduction}

The research on ultracold atoms has been significantly enriched by
the realization of synthetic spin-orbit coupling (SOC), i.e., of induced
interaction between the center-of-mass momentum and internal (hyperfine)
states \cite{key-SpielmanNature1,key-SpielmanPRA}. The development
of different variations of the basic scenario, with the application
to, both, bosonic and fermionic systems \cite{key-fermionSOC1,key-fermionSOC2},
has paved the way to an active area of research, where the advances
in theory and experiments are continuous. A variety of fundamental
effects have been uncovered and powerful technical implications of
the findings are expected. Particularly interesting is the potential
emergence in these systems of novel states of matter, like nontrivial
superfluids or topological insulators \cite{key-SpielmanNature1,key-SpielmanRepProg,key-DalibardRMP}.
The implementation of strategies for controlling different aspects
of the dynamics is crucial for the advances in this line \cite{key-SpielmanControl,key-ZhangSciRep,key-Llorente}.
Here, we focus on recent experiments on spin-orbit coupled Bose-Einstein
condensates which uncovered the possibility of using the external
dynamics to manipulate the spin polarization \cite{key-lzSOC}. Specifically,
in a Raman-induced SOC setup \cite{key-lzSOC}, acceleration of the
external modes due to gravitational or harmonic trapping forces was
shown to induce transitions between energy bands associated with eigenvalues
of the internal-state reduced Hamiltonian. In a preliminary analysis
of the results \cite{key-lzSOC}, the detected features were found
to be reproduced by the Landau-Zener (LZ) model. In particular, the
measured probabilities of transition between spin states agreed with
the predictions of the model in the asymptotic limit. In this sense,
the study has a certain predictive power that allows the possibility
of tuning the transition characteristics. Still, the explanation of
some aspects of the physical mechanisms responsible for the observed
behavior requires additional work. Namely, a more complete characterization
of the emergence of the LZ model in those systems is needed. The role
of the external dynamics in the inter-band transitions must be clarified.
Also important is to explain the observed robustness of the model
against variations in the characteristics of the acceleration methods.
It is worth pointing out that, although the terms added to the basic
SOC Hamiltonian in the two considered schemes have different functional
forms, the global output is similar. An additional open question refers
to the relevance of many-body effects to the transition processes.
To deal with those issues, we will analytically study the two implemented
setups. In our analysis, the applicability of the LZ model will be
rigorously traced and the origin of the parallelism observed between
the two schemes will be identified. Moreover, effects specific to
the properties of the initial state will be tackled: we will see that,
for the LZ approach to be applicable, restrictions on the system preparation
must be imposed. We will also assess the robust character of the single-particle
approach against atomic-interaction effects.

The outline of the paper is as follows. In Sec. II, the main characteristics
of the system in the absence of acceleration schemes are summarized.
Additionally, we discuss how the internal dynamics is altered when
acceleration of the external modes is introduced in the basic scenario.
The relevance of the LZ model to the description of different variations
of the system is evaluated. In Sec. III, the effect of the gravitational
field is studied. Through an appropriate sequence of unitary transformations
and an ultimate reduction to the spin space, the Hamiltonian of the
complete system is cast into the basic form of the LZ model. Some
aspects specific to the preparation of the system and the measurement
procedure are discussed. Sec. IV is dedicated to the analysis of the
acceleration due to harmonic trapping forces. A method based on the
eikonal approximation is applied first to put the focus on the dynamics
of the transitions. Subsequently, an approximate analytical study
of the evolution outside the crossing region is given to complete
the picture. The relevance of many-body effects to the applicability
of our approach is discussed in Sec. V.  Finally, some general conclusions
are summarized in Sec. VI.

\section{The Landau-Zener model in spin-orbit coupled atoms}

The inter-band transitions on which we focus can take place in different
realizations of SOC in Bose-Einstein condensates \cite{key-SpielmanRepProg}.
Here, without loss of generality, we will base our analysis on the
setup presented in Ref. \cite{key-SpielmanNature1}. There, synthetic
SOC was generated via an appropriate arrangement of Raman lasers.
Specifically, one of the components of the external linear momentum
of the atoms $\mathbf{p}$ was coupled to an effective ``spin'',
i.e., to a two-level internal system formed by hyperfine states. The
arrangement incorporated two orthogonally polarized Raman lasers with
different propagation directions and frequencies. The setup was configured
to couple two Zeeman-split internal states; furthermore, the coupling
was made to be dependent on the atom external momentum. In this form,
an effective SOC was created. Actually, in the considered experimental
setup \cite{key-lzSOC}, as in previous standard realizations of synthetic
SOC, the three states $\left|m_{F}\right\rangle $ of the hyperfine
$F=1$ ground-state manifold of $^{87}\mathrm{Rb}$ are coupled. It
is the detuning due quadratic Zeeman shift of the state with $m_{F}=1$
that allows the reduction of the dynamics to the two levels with $m_{F}=0,-1$.
A first picture of the dynamics is provided by a basic model where
a single-particle description is applied and the effect of the harmonic
confinement is ignored. Intense work on different implementations
of SOC, which include variations in the coupling and dimensionality,
is being carried out. Here, as in the experimental realization of
\cite{key-lzSOC}, we will focus on the setup corresponding to Rashba-Dresselhaus
SOC. In the rest of the paper, the term SOC will refer exclusively
to that type of coupling. The associated Hamiltonian is given by 

\begin{equation}
H_{SO}=\frac{P_{x}^{2}}{2m}+\frac{\hbar\Omega}{2}\sigma_{x}+\left(\frac{\hbar\delta_{0}}{2}+\frac{\alpha_{0}P_{x}}{\hbar}\right)\sigma_{z}.\label{eq:SOCHamiltonian}
\end{equation}
The atomic mass is denoted by $m$, $\sigma_{i}$ ($i=x,\, y\,,z$)
are the Pauli matrices corresponding to the \emph{pseudospin}, i.e.,
to the considered effective two-level system, and $P_{x}$ stands
for the momentum operator in the coupling direction. The additional
parameters refer to the laser characteristics: $\Omega$, usually
termed as the \emph{offset}, represents the Raman coupling amplitude,
$\delta_{0}$ denotes an adjustable detuning, and $\alpha_{0}$ is
the strength of the realized SOC ($\alpha_{0}=2E_{r}/k_{r}$ where
$E_{r}=\hbar^{2}k_{r}^{2}/2m$ and $k_{r}=2\pi\sin(\theta/2)/\lambda$
are, respectively, the recoil energy and momentum; $\lambda$ is the
reference laser wavelength, and, $\theta$ is the angle between the
directions of the Raman lasers, which, here, as in the experiments
in \cite{key-SpielmanNature1,key-SpielmanPRA}, is fixed to $\theta=\pi/2$). 

Since $P_{x}$ is a constant of the motion, it is convenient to work
in the basis $\left|p_{x}\right\rangle \otimes\left|\chi\right\rangle $,
where $\left|p_{x}\right\rangle $ denotes a momentum state and $\left|\chi\right\rangle $
stands for a spin state. The consequent reduction of the Hamiltonian
to the spin space is simply obtained by replacing the operator $P_{x}$
by $\hbar k_{x}$ in Eq. (\ref{eq:SOCHamiltonian}). ($k_{x}$ is
the quasi-momentum in the coupling direction, $\mathbf{p}=\hbar\mathbf{k}$).
Using this approach, the system is found to present energy bands given
by 

\begin{equation}
E_{\pm}(k_{x})=\frac{\hbar^{2}k_{x}^{2}}{2m}\pm\sqrt{\left(\frac{\hbar\Omega}{2}\right)^{2}+\left(\frac{\hbar\delta_{0}}{2}+\alpha_{0}k_{x}\right)^{2}}.\label{eq:bands}
\end{equation}
Since the bands correspond to eigenvalues of the reduced Hamiltonian,
no coupling between them exists in this scheme. They reflect the existence
of locking between the momentum and the spin state: each quasi-momentum
value is attached to a particular eigenvalue, and, in turn, to the
corresponding combination of internal states. We will see that inter-band
transfers of population can take place in variations of the above
setup, specifically, in the arrangements implemented in Ref. \cite{key-lzSOC}.
It will be shown that the transitions can be described in terms of
switching between adiabatic states in the LZ model \cite{key-Landau,key-Zener}.
Actually, the adiabatic regime can be already identified: an external
mechanism that can induce a slow variation of $k_{x}$ can be expected
to lead to an adiabatic following of the internal state along the
band, the \emph{static} locking being trivially conserved in the resulting
evolution. 

Let us recall some basic characteristics of the LZ approach. In its
simplest form, this model deals with a system of two coupled states
(the diabatic states) with a linear variation of the energy mismatch.
The associated Hamiltonian reads 

\begin{equation}
H_{1}=\hbar\frac{vt}{2}\sigma_{z}+\hbar\zeta\sigma_{x},\label{eq:LZmodel}
\end{equation}
where $v$ denotes the rate of the splitting variation, and $\zeta$
represents the coupling strength. An alternative description in terms
of the adiabatic states, i.e., of the instantaneous eigenstates of
the coupled system, is also convenient. Both pictures give the same
information on the dynamics. We will switch from one description to
the other as it can be demanded by the clarification of the physics
underlying the studied transitions. At the initial and final times,
which are formally defined as $t\rightarrow\mp\infty$, the energy
levels are assumed to be far apart. Therefore, at those times, the
adiabatic states can be considered to approximately match the diabatic
states. Actually, $\zeta$ is assumed to be sufficiently small  for
the coupling to be effective only near the crossing that takes place
between the diabatic levels at $t=0$. In those conditions, the probability
of transition between diabatic states at the end of the linear ramp
is given by 
\begin{equation}
P_{LZ}^{(d)}=1-e^{-2\pi\left|\zeta\right|^{2}/v}.
\end{equation}
The counterpart description in the adiabatic basis depicts an avoided
crossing at $t=0$, the asymptotic probability of transition between
adiabatic states being given by $P_{LZ}^{(a)}=e^{-2\pi\left|\zeta\right|^{2}/v}$.
Note that the adiabatic regime, characterized by $P_{LZ}^{(a)}=0$,
is approached as the ratio $\left|\zeta\right|^{2}/v$ is increased.
The LZ model also applies to variations of the basic scenario described
by Eq. (\ref{eq:LZmodel}) which incorporate an independent term in
the linearly-modified energy mismatch, i.e., to Hamiltonians with
the form

\begin{equation}
H_{2}=\hbar(\Lambda+\frac{vt}{2})\sigma_{z}+\hbar\zeta\sigma_{x}.\label{eq:LZmodel-1}
\end{equation}
Moreover, the initial and final times of the process do not need to
correspond to $t\rightarrow\mp\infty$. The only restrictions are
the existence of a crossing, i.e., the occurrence of a zero value
of the energy mismatch during the ramping, and the need of having
sufficiently large values of the magnitude of the initial and final
splittings. Relevant to the cases that will be analyzed further on
in this paper is to refer to an initial time $t_{i}=0$, and, consequently
to an initial splitting given by the independent term $\hbar\Lambda$.
It is apparent that a positive (negative) $\Lambda$ requires a negative
(positive) ramp rate $v$ for the restrictions of the LZ model to
be fulfilled. The probability of transition for the case of negative
rate is obtained by simply changing $v$ by $\left|v\right|$ in Eq.
(4). Hence, in the following, to account for both cases, we will respectively
use the expressions $P_{LZ}^{(d)}=1-e^{-2\pi\left|\zeta\right|^{2}/\left|v\right|}$
and $P_{LZ}^{(a)}=e^{-2\pi\left|\zeta\right|^{2}/\left|v\right|}$
for the diabatic and adiabatic probabilities of transition. 

Let us evaluate the applicability of the LZ approach to the SOC setup.
In the system considered by now, see Eq. (\ref{eq:SOCHamiltonian}),
the splitting between the spin states is fixed since $P_{x}$ is a
constant of the motion ($k_{x}$ enters the reduced description as
a fixed parameter). In fact, a level crossing occurs only for a specific
value of the quasi-momentum $k_{x,c}$, namely, for that given by
the condition

\begin{equation}
k_{x,c}=-\frac{\hbar\delta_{0}}{2\alpha_{0}}.\label{eq:MomentCrossVal}
\end{equation}
Hence, $k_{x,c}$ is determined by the detuning and by the SOC strength;
in particular, $k_{x,c}=0$, for zero detuning. Obviously, no analogy
with the LZ model can be traced at this point. However, we can think
of variations of the above \emph{static} scenario where a modification
of the spin-state splitting can be implemented via the control of
the external dynamics. Specifically, one can propose setups where
the acceleration of the system in the $x$-direction can be used to
vary the energy mismatch. This is the case of the situations considered
in the experiments reported in \cite{key-lzSOC}: the acceleration
mechanisms were incorporated to drive the momentum to the range of
values where $k_{x}\simeq k_{x,c}$. There, the coupling incorporated
in the offset becomes effective to activate the crossing, and, consequently,
 transitions between internal states can be induced. Here, the applicability
of the LZ model can be reasonably conjectured. The dispersion curves
(\emph{dressed bands}) given by Eq. (\ref{eq:bands}) can be thought
of corresponding to the adiabatic levels of the LZ model. The existence
of an avoided level crossing is apparent in them; the magnitude of
the associated gap is given by the \emph{offset}: as $\Omega$ increases,
the adiabatic bands progressively separate. Note that, in the adiabatic
picture, it is an inter-band transition that is brought about by the
acceleration. The diabatic counterparts  (\emph{bare bands}) correspond
to taking $\Omega=0$ in those curves: the offset disappears from
the expressions for the energy levels and incorporates the coupling
between (diabatic) spin states. Whereas, in the primary LZ setup,
it is a driving term, independent of the considered dynamics, that
induces the linear variation of the energy mismatch, in our system,
it is the dynamics of the external variables that leads to the modification
of the internal-state splitting. In the next sections, we will analytically
trace the occurrence of LZ transitions generated by gravitational
and harmonic-trapping forces. Dealing with the required enlarged dimensionality
of the problem is one of the aims of our work. Another basic objective
is to analyze the implications to the extended system of the conditions
of having large initial and final splittings, required for the applicability
of the basic LZ model.

\section{Gravitational acceleration}

Let us analyze first the case where the acceleration of the system
is induced by the gravitational field. In the experiments \cite{key-lzSOC},
that situation was arranged by implementing the SOC in the vertical
axis. Then, the trap was turned off, and the system, prepared as a
Gaussian superposition of momentum states, was led to evolve under
the effect of the gravity. Accordingly, we consider here that, at
$t=0$, the gravitational field is \emph{connected}, the dynamics
being then governed by the Hamiltonian

\begin{equation}
H_{G}=H_{SO}+H_{grav},\label{eq:gHamilt}
\end{equation}
where

\begin{equation}
H_{grav}=mgX.
\end{equation}
Note that, in our theoretical framework, which corresponds to non-relativistic
quantum mechanics, the effect of gravity can be regarded as that of
an effective external electric field. As shown in \cite{key-kasevich},
this approach is convenient to operatively account for nontrivial
effects of gravity on a quantum context. 

In the setup provided by Eqs. (7) and (8), $P_{x}$ is not a constant
of the motion. Indeed, as previously stated, the variation of the
momentum, i.e., the acceleration of the system, is the objective of
incorporating $H_{grav}$. Appropriate to analyze the dynamics is
the application of the unitary transformation

\begin{equation}
U_{1}(t)=\exp\left[-\frac{i}{\hbar}mgXt\right],\label{eq:U1g}
\end{equation}
which introduces a time-dependent displacement in momentum, specifically,
it implies working in a reference frame translated with acceleration
$g$. The transformed Hamiltonian, given by 

\begin{equation}
H_{G}^{\prime}=U_{1}^{\dagger}H_{G}U_{1}-i\hbar U_{1}^{\dagger}\dot{U}_{1},
\end{equation}
is written, after straightforward algebra \cite{key-Louisell}, as

\begin{eqnarray}
H_{G}^{\prime} & = & \frac{P_{x}^{2}}{2m}-gP_{x}t+\frac{\hbar\Omega}{2}\sigma_{x}+\nonumber \\
 &  & \frac{\hbar\delta_{0}}{2}\sigma_{z}+\frac{\alpha_{0}}{\hbar}\left(P_{x}-mgt\right)\sigma_{z},\label{eq:HprimeG}
\end{eqnarray}
where we have shifted the energy origin by $mg^{2}t^{2}/2$. Now,
a second unitary transformation with the form 

\begin{equation}
U_{2}(t)=\exp\left[-\frac{i}{\hbar}\frac{P_{x}^{2}}{2m}t+\frac{i}{\hbar}gP_{x}\frac{t^{2}}{2}\right]\label{eq:gaugetransf}
\end{equation}
is applied to incorporate the spin-independent terms present in Eq.
(\ref{eq:HprimeG}). $U_{2}(t)$ parallels a gauge transformation
which simply introduces a phase depending on both, momentum and time
\cite{key-Sakurai}. We obtain for the transformed Hamiltonian

\begin{eqnarray}
H_{G}^{\prime\prime} & = & U_{2}^{\dagger}H_{G}^{\prime}U_{2}-i\hbar U_{2}^{\dagger}\dot{U}_{2}\nonumber \\
 & = & \frac{\hbar\Omega}{2}\sigma_{x}+\left[\frac{\hbar\delta_{0}}{2}+\frac{\alpha_{0}}{\hbar}\left(P_{x}-mgt\right)\right]\sigma_{z}.
\end{eqnarray}
Again working in the representation $\left|p_{x}\right\rangle \otimes\left|\chi\right\rangle $,
we can write for the reduction of the Hamiltonian to the spin space
the expression 
\begin{equation}
H_{G}^{\prime\prime}=\frac{\hbar\Omega}{2}\sigma_{x}+\left[\frac{\hbar\delta_{0}}{2}+\frac{\alpha_{0}}{\hbar}\left(\hbar k_{x}-mgt\right)\right]\sigma_{z},\label{eq:LZg}
\end{equation}
where the quasi-momentum plays the role of a parameter. The analogy
with the LZ model is now apparent: the resulting effective two-level
system, coupled through the offset term, displays a linear variation
of the energy mismatch. For a state with quasi-momentum $k_{x}$,
the crossing time $t_{c}$ between the diabatic states is given by
$t_{c}=\frac{\hbar\delta_{0}/2+\alpha_{0}k_{x}}{\alpha_{0}mg/\hbar}$;
therefore, it depends on both, $k_{x}$ and $\delta_{0}$. Importantly,
in order to apply the restriction of having initially a large splitting,
required for applying the LZ model, we must take into account that,
here, the initial time is $t=0$. The fulfillment of that restriction
is then guaranteed by working with a sufficiently large (positive)
quasi-momentum. The comparison with Eq. (\ref{eq:LZmodel}) leads
to the identification of the LZ characteristic parameters as

\begin{eqnarray}
\zeta & \rightarrow & \frac{\Omega}{2}\nonumber \\
v & \rightarrow & -\frac{2\alpha_{0}mg}{\hbar^{2}}
\end{eqnarray}
Hence, the coupling strength is given by the offset, and the rate
of splitting variation is determined by the SOC amplitude and by the
atom weight. The presence of the detuning $\delta_{0}$ and the quasi-momentum
$k_{x}$ in Eq. (\ref{eq:LZg}) does not affect the applicability
of the model. Although they alter the value of the crossing time $t_{c}$,
they do not affect the asymptotic transition probability, which is
still determined by the characteristic LZ parameters, provided that
the splittings corresponding to the initial and final times are sufficiently
large. Figure 1 illustrates these results.

\begin{figure}[H]
\centerline{\includegraphics{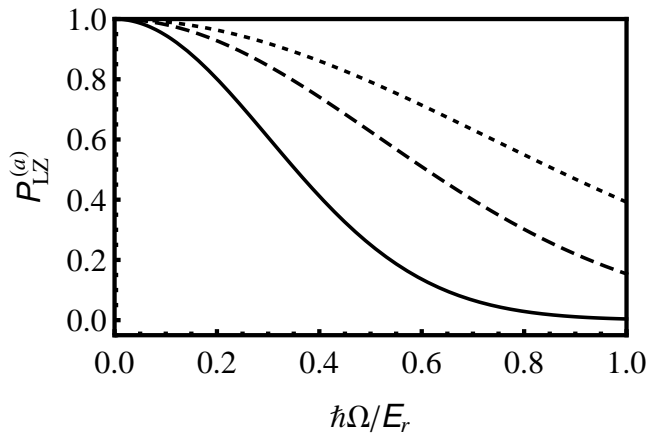}}\caption{The probability of transition between adiabatic states as a function
of the offset parameter (in units of the recoil energy, $E_{r}/2\pi\hbar=3.75$
kHz). The continuous line corresponds to gravitational acceleration;
The dashed line refers to harmonic acceleration with $\omega_{x}/2\pi=70$
Hz and $q_{0}=1.5\times10^{-4}$ m; The dotted line represents harmonic
acceleration with $\omega_{x}/2\pi=70$ Hz and $q_{0}=3.\times10^{-4}$
m. }
\end{figure}

The above results can be applied to explain the experimental findings.
The practical realization corresponded to a Gaussian superposition
of momentum states. Moreover, it was the probability of transition
between spin states, independently on the momentum values, that was
measured. Hence, in a description in terms of the density operator
of the complete system, (including, both, internal and motional variables),
the experimental procedure formally corresponds to carrying out a
partial trace over the momentum states. Actually, given the characteristics
of the measurement procedure, it is not a pure state but a statistical
mixture that is required to simulate the experimental realization.
Therefore, the distribution of momentum populations of the initial
preparation contains all the information necessary to reproduce the
measured transition probabilities. The distribution corresponding
to the experiments was assumed to be well simulated by the function
\begin{equation}
\rho(k_{x})=\frac{1}{\sqrt{2\pi}\sigma_{w}}\exp\left[-(k_{x}-k_{x,i})^{2}/(2\sigma_{w}^{2})\right],\label{eq:MomDistr}
\end{equation}
where the center $k_{x,i}$ and the width $\sigma_{w}$ of the Gaussian
are parameters specific to each realization. The total probability
of transition is obtained by averaging, over the distribution, the
partial probabilities for the different momenta, all of them given
by the LZ formula $P_{LZ}^{(d)}=1-e^{-2\pi\left|\zeta\right|^{2}/\left|v\right|}$.
In the assumed conditions, although there is a variety of crossing
times, there is a unique value of the asymptotic transition probability.
We recall that no amplitudes of transition are evaluated. The parameters
$k_{x,i}$ and $\sigma_{w}$ must be chosen in order to guarantee
that the initial splittings, for all the states that contribute to
the wave-packet, are sufficiently large. Moreover, the measurement
must be carried out when the momenta have reached large-magnitude
values. Consequently, the system can be assumed to be far from the
diabatic crossing at the initial and final times, as required for
applying the \emph{standard} LZ approach. These restrictions are expressed
as 
\begin{equation}
\left|\frac{\hbar\delta_{0}}{2}+\alpha_{0}\left(k_{x,i}\pm\sigma_{w}\right)\right|\gg\frac{\hbar\Omega}{2},\label{eq:restricLZ}
\end{equation}
for the parameters of the initial distribution, a similar equation
being applicable to the final momentum values. The above expression
has been obtained taking into account that the crossing region corresponds
to an energy mismatch comparable to the magnitude of the coupling
term $\frac{\hbar\Omega}{2}$. Note the importance of the magnitude
of $\Omega$ in validity of the model. The relevance of the SOC parameter
$\alpha_{0}$ must also be pointed out. For a direct comparison with
the parameters used in the characterization of the experimental setup,
we must recall the dependence of $\alpha_{0}$ on the recoil energy
and momentum, namely, $\alpha_{0}=2E_{r}/k_{r}$. Note that as $\alpha_{0}$
decreases, the magnitude of the SOC diminishes, and, eventually, the
effect vanishes. Then, a significant value of $\alpha_{0}$ is needed
for our characterization of the system to be meaningful. One should
also take into account that $\delta_{0}$ affects the location of
the crossing point; additionally, it enters Eq. (17), which guarantees
the validity of the LZ model, i.e., the accomplishment of the asymptotic
regime in both initial and final configurations. We will go back to
this point in different parts of the paper given the importance of
the definition of the asymptotic regimes. In all the experimental
realizations of \cite{key-lzSOC}, those restrictions are amply fulfilled.

In passing, our analysis provides an understanding for one conspicuous
feature detected in the experiments. Namely, the observed lack of
dependence of the probability of transition on the detuning is explained
by the mere applicability of the LZ model: the detuning and the initial
values of $k_{x}$, are irrelevant to the model predictions for the
asymptotic probabilities. 

Two additional remarks are pertinent. First, we have assumed that
the diabatic states can be exactly identified with the internal states
of the SOC Hamiltonian. In fact, in the experiments, the initial state
corresponded to a combination of the two internal states with one
of the associated coefficients having a magnitude much larger than
the other. Our approach is still applicable to that situation: the
output of the process is then a combination of states where the squared
moduli of the coefficients have asymptotic values given by the predictions
of the LZ for the counterpart probabilities of transition. Second,
the effective rate of splitting variation $v$ is fixed since it is
determined by the gravitational acceleration and by the SOC strength
$\alpha_{0}$. Therefore, there is no form of reaching an adiabatic
evolution by reducing $v$. The only nontrivial possibility of realizing
an effective adiabatic regime, i.e., of diminishing $P_{LZ}^{(a)}=e^{-2\pi\left|\zeta\right|^{2}/\left|v\right|}$,
is by increasing the offset. We stress that the enhancement in $\Omega$
needed to reach the adiabatic regime can be made to be compatible
with the restriction given by Eq. (\ref{eq:restricLZ}). Indeed, this
is the case of the experimental realization of the adiabatic regime
\cite{key-lzSOC}.

\section{Harmonic acceleration}

Let us consider now the case where the acceleration of the system
is induced by the trapping potential. Accordingly, the dynamics is
assumed to be governed by the Hamiltonian 

\begin{equation}
H_{T}=H_{SO}+H_{trap}\label{eq:SOCharmonic}
\end{equation}
where 

\begin{equation}
H_{trap}=\frac{1}{2}m\omega_{x}^{2}X^{2}.
\end{equation}
Again, $P_{x}$ is not a constant of the motion. One can anticipate
that, for the system prepared in a non-stationary state of the complete
Hamiltonian, the trap provides an acceleration mechanism, which changes
the mean value of $P_{x}$, and, consequently, the splitting between
spin states.  In practice, the initial out-of-equilibrium situation
was realized through the gravitational acceleration, i.e., via the
scheme analyzed in the previous section: the system was prepared in
a Gaussian superposition of momentum states whose center, $k_{x,i}$,
was arranged to have a large-magnitude value, see Eq. (\ref{eq:MomDistr}).
We intend to characterize the evolution followed by that state when
the trap is connected. We will evaluate the conditions required for
driving the momentum to the range of values where a transition between
spin states becomes feasible. Moreover, we will identify the origin
of the parallelism observed with the transitions induced by the gravitational
scheme.

\subsection{Description of the inter-band transitions through the eikonal approximation}

As opposed to the case of the gravitational field, a complete analytical
treatment of the system dynamics is not possible for a harmonic force.
Still, some approximations can be implemented to analytically trace
the applicability of the LZ model. Convenient for this aim are some
considerations on the classical dynamics of the motional variables
in the trap. Previous to the transition and provided that the coupling
effects are neglected outside the crossing, the evolution of the system
corresponds to a harmonic oscillator. Hence, the classical trajectory
is characterized by a sinusoidal variation of both the coordinate
and the momentum. From this classical picture, one can predict that,
for the initial conditions corresponding to the considered out-of-equilibrium
situation, the momentum eventually reaches the (reduced) value necessary
for a crossing between spin states to take place. We recall that the
momentum at the crossing, $k_{x,c}$, is given by Eq. (\ref{eq:MomentCrossVal});
in particular, as previously pointed out, one finds $k_{x,c}=0$ for
zero detuning. The required significant reduction in momentum implies
that the initial energy has been converted, in large extent, into
potential energy at the crossing. Additionally, we can expect that,
in the analysis of the quantum dynamics, in order to put the focus
on the transition region, a unitary transformation that incorporates
 a spatial translation to the crossing must be appropriate. Accordingly,
we apply 

\begin{equation}
U=\exp\left[-\frac{i}{\hbar}q_{0}P_{x}\right].\label{eq:eikonalUnitTransf}
\end{equation}
where the (static) displacement  $q_{0}$ is chosen as the magnitude
of the position at the crossing, i.e., of the coordinate corresponding
to $k_{x}=k_{x,c}$. Following the previous considerations on the
classical dynamics, we evaluate $q_{0}$ through the energy conservation
\begin{equation}
q_{0}=\sqrt{\frac{2}{m\omega_{x}^{2}}}\left[\left\langle \frac{P_{x}^{2}}{2m}+\frac{\alpha_{0}P_{x}}{\hbar}+\frac{1}{2}m\omega_{x}^{2}X^{2}\right\rangle _{initial}-\left(\frac{(\hbar k_{x,c})^{2}}{2m}+\alpha_{0}k_{x,c}\right)\right]^{1/2},\label{eq:q0value}
\end{equation}
where $\left\langle \right\rangle _{initial}$ denotes an average
over the initial state. Note that $q_{0}$ increases with the energy.
The transformed Hamiltonian is given by

\begin{equation}
H_{T}^{\prime}=U^{\dagger}H_{T}U=\frac{P_{x}^{2}}{2m}+\frac{1}{2}m\omega_{x}^{2}(X+q_{0})^{2}+\frac{\hbar\Omega_{0}}{2}\sigma_{x}+\left(\frac{\hbar\delta_{0}}{2}+\frac{\alpha_{0}P_{x}}{\hbar}\right)\sigma_{z},
\end{equation}
where, the term $\frac{1}{2}m\omega_{x}^{2}X^{2}$ can be considered
to be much smaller than $m\omega_{x}^{2}q_{0}X$ for a large enough
displacement $q_{0}$, i.e., for an initial preparation corresponding
to a sufficiently high energy. Then, $H_{T}^{\prime}$ can be approximated
as 

\begin{equation}
H_{T}^{\prime}=\frac{P_{x}^{2}}{2m}+m\omega_{x}^{2}q_{0}X+\frac{\hbar\Omega_{0}}{2}\sigma_{x}+\left(\frac{\hbar\delta_{0}}{2}+\frac{\alpha_{0}P_{x}}{\hbar}\right)\sigma_{z},\label{eq:harHamRedg}
\end{equation}
where we have shifted the energy origin by $\frac{1}{2}m\omega_{x}^{2}q_{0}^{2}$.
We emphasize that the restriction $\left\langle \frac{1}{2}m\omega_{x}^{2}X^{2}\right\rangle \ll\left\langle m\omega_{x}^{2}q_{0}X\right\rangle $,
required by the validity of our derivation is guaranteed by the fulfillment
of Eq. (\ref{eq:restricLZ}) for both, initial and final, momentum
values. This is straightforwardly shown using the value of $q_{0}$,
given by Eq. (\ref{eq:q0value}) and estimating the values of $\left\langle X\right\rangle $
in the crossing region, e.g. via energy conservation. It is important
to stress that no additional limitations are being introduced in our
framework.

Our approximate description of the system can be regarded as resulting
from the application of the eikonal approximation \cite{key-Sakurai}.
Namely, in our procedure, the use of the unitary transformation that
incorporates the displacement to the crossing point is equivalent
to propose an ansatz for the complete wave function in terms of the
product of a phase term and a reduced wave function in the momentum
representation as the product of a highly oscillatory phase and a
reduced smooth wave function. Specifically, we employ a phase linearly
dependent on the momentum with a slope to be determined. Then, the
optimum value of the extracted phase  is obtained by requiring it
to simplify the reduced Schrödinger equation. The applied simplification
of the Hamiltonian strongly relies on the elimination of the term
of harmonic confinement. For it to be sound, a large magnitude for
the displacement is required. Moreover, the validity of the approach
is improved with the accuracy of the translation to the crossing,
i.e., with the precision in the determination of $q_{0}$. A more
elaborate derivation of $q_{0}$, in a complete quantum formalism
which incorporates details of the practical implementation, will be
presented further on. The simple procedure followed by now is intended
to stress the way in which the energy-dependent extracted phase affects
the reduced Hamiltonian, and, in turn, as we shall see, the role of
the energy of the initially prepared state in the characterization
of the effective LZ parameters.

It is apparent that the Hamiltonian in Eq. (\ref{eq:harHamRedg})
has the same form as that of the gravitational acceleration, given
by Eq. (\ref{eq:gHamilt}). Here, the term $m\omega_{x}^{2}q_{0}X$
is the counterpart of the gravitational potential $H_{grav}=mgX$.
The emergence of the LZ model in this scenario can be directly traced,
as in the previous section,  by consecutively applying two unitary
transformations similar to those given by Eqs. (\ref{eq:U1g}) and
(\ref{eq:gaugetransf}). Accordingly, and, after the reduction to
the spin space, we find 
\begin{equation}
H_{T}^{\prime\prime}=\frac{\hbar\Omega}{2}\sigma_{x}+\left[\frac{\hbar\delta_{0}}{2}+\frac{\alpha_{0}}{\hbar}\left(\hbar k_{x}-m\omega_{x}^{2}q_{0}t\right)\right]\sigma_{z}.
\end{equation}
Now, the comparison with Eq. (\ref{eq:LZmodel}) leads to the identification
of the LZ characteristic parameters as:

\begin{eqnarray}
\zeta & \rightarrow & \frac{\Omega}{2}\nonumber \\
v & \rightarrow & -\frac{2\alpha_{0}m\omega_{x}^{2}q_{0}}{\hbar^{2}}
\end{eqnarray}
As in the gravitational scheme, the coupling strength is given by
the offset. Common to the two considered acceleration schemes is also
the proportionality existent between the effective splitting-variation
rate $v$ and the SOC strength $\alpha_{0}$. Yet, there is a differential
effect: in the harmonic setup, as opposed to the behavior found in
the gravitational scheme, $v$ depends on the characteristics of the
system preparation, in particular, on the energy. That dependence
is incorporated in Eq. (25) through the trap frequency $\omega_{x}$
and the displacement to the crossing $q_{0}$. The magnitude of $v$
increases with $q_{0}$, and, therefore, with the energy of the system.
Figure 1 illustrates the analogies and differences between the two
considered acceleration schemes. The used parameters are similar to
those corresponding to the experimental realization \cite{key-lzSOC}. 

The simplified picture of the dynamics given by our approach uncovers
the components of the system that are actually relevant to the LZ
transition. It is the rate of variation of the momentum at the intersection
point that determines the probability of switching between spin states.
In turn, that rate is fixed by the energy of the initial preparation
as far as transitions outside the crossing region can be neglected.
As in the gravitational-acceleration scheme, it is found that the
detuning does not affect the transition probability. Importantly,
for our derivation to be valid, the preparation of the system must
correspond to a sufficiently high energy. There are limitations on
the energy values that can be achieved in the practical arrangement.
Still, let us remind again that the conditions contained in Eq. (17)
are safely satisfied by the experiments, which explains the robustness
of the LZ approach: despite the variety of used parameters, the model
was found to accurately reproduce the observed features. It is important
to point out that, in our analysis, as in the experiments, only one
crossing is being considered. The periodic character of the dynamics
of the external variables in the harmonic trap is not tackled.

\subsection{A model for the dynamics outside the crossing}

In the former analysis, the focus was put on the transition region.
The optimum displacement $q_{0}$ used in the unitary transformation
given by Eq. (\ref{eq:eikonalUnitTransf}) was evaluated through simple
classical arguments on energy conservation. Let us present now a more
complete (quantum) picture of the dynamics previous to the crossing
which incorporates specific aspects of the experimental realization.
This framework will provide us with another method for deriving $q_{0}$.

In order to characterize the evolution of the diabatic states, we
can reasonably assume, as previously, that the coupling, accounted
for by the offset term in the Hamiltonian, becomes effective only
very near the crossing. Hence, as the transition is approached, the
evolution of the initially prepared state can be regarded as given
by the Hamiltonian in Eq. (\ref{eq:SOCharmonic}) where the offset
is neglected in the term $H_{SO}$. The resulting Hamiltonian becomes
spin-separable, and, for each spin value, it corresponds to a displaced
harmonic oscillator. Namely,  the dynamics before the crossing can
be assumed to be governed by 

\begin{equation}
H_{T,\pm}=\hbar\omega_{x}a^{\dagger}a\pm\left(\frac{\hbar\delta_{0}}{2}+\frac{\alpha_{0}}{\hbar}i\sqrt{\frac{m\hbar\omega_{x}}{2}}(a^{\dagger}-a)\right),
\end{equation}
where $\pm$ refers to the spin components, and, $a^{\dagger}$ and
$a$ are the creation and annihilation operators associated to the
$x$-coordinate.

The procedure to obtain the eigenstates of $H_{T,\pm}$ is straightforward.
Let us illustrate here its application to $H_{T,+}.$ First, we use
the unitary transformation given by 

\begin{equation}
U=D(\beta),
\end{equation}
where $D(\beta)$ is the displacement operator \cite{key-Louisell}
with argument $\beta$ to be determined by imposing the intended reduction
of the description. The transformed Hamiltonian is given by 

\begin{equation}
H_{T,+}^{\prime}=D^{\dagger}(\beta)H_{T,+}D(\beta)=\hbar\omega_{x}(a^{\dagger}+\beta^{\star})(a+\beta)+\frac{\hbar\delta_{0}}{2}+\frac{\alpha_{0}}{\hbar}i\sqrt{\frac{m\hbar\omega_{x}}{2}}(a^{\dagger}+\beta^{\star}-a-\beta).
\end{equation}
Now, by choosing 

\begin{equation}
\beta=-i\frac{\alpha_{0}}{\hbar}\sqrt{\frac{m}{2\hbar\omega_{x}}},
\end{equation}
$H_{T,+}^{\prime}$ is converted into the Hamiltonian of an undisplaced
harmonic oscillator, namely, into

\begin{equation}
H_{T,+}^{\prime}=\hbar\omega_{x}a^{\dagger}a+\hbar\omega_{x}\left|\beta\right|^{2}-\frac{\alpha_{0}^{2}}{\hbar^{2}}m,
\end{equation}
where the constant terms simply shift the energy origin. Important
for the comparison with the former derivation of $q_{0}$ is to take
into account that the applied unitary transformation implies working
with a displaced momentum, the associated operator $\tilde{P_{x}}=D^{\dagger}(\beta)P_{x}D(\beta)$
being given by

\begin{equation}
\tilde{P_{x}}=P_{x}+\sqrt{2m\hbar\omega_{x}}\textrm{Im}(\beta)=P_{x}+\frac{\alpha_{0}}{\hbar}m.
\end{equation}
It is pertinent to refer here to the role of the gravitational acceleration
in the current scheme, where the harmonic trap is connected. Actually,
the gravitational potential can be included in our framework through
a unitary transformation that displaces the harmonic oscillator in
the vertical coordinate. Therefore, it does not introduce qualitative
differences in the dynamics: it simply modifies the equilibrium position
of the trapping potential.

In the representation of eigenstates of $H_{T+}^{\prime}$, the evolution
of the initially prepared state is direct. This is the case, in particular,
if we assume that the initial preparation corresponds to a coherent
state, $\left|\eta\right\rangle $, i.e., to an eigenstate of the
annihilation operator, $a\left|\eta\right\rangle =\eta\left|\eta\right\rangle $,
where $\eta$ is the complex number from which all the state characteristics
are determined. Hence, we take for the initial state

\begin{equation}
\left|\varphi(t=0)\right\rangle =\left|\eta\right\rangle ,
\end{equation}
whose associated wave function is given by \cite{key-CohenBookQuantumM}
\begin{equation}
\varphi(x,t=0)=\left\langle x\right.\left|\eta\right\rangle =e^{i\theta_{\eta}}\left(\frac{m\omega_{x}}{\pi\hbar}\right)^{1/4}\exp\left\{ -\left[\frac{x-\left\langle X\right\rangle _{\eta}}{2\Delta X_{\eta}}\right]^{2}+i\left\langle P\right\rangle _{\eta}\frac{x}{\hbar}\right\} 
\end{equation}
with
\begin{eqnarray}
\left\langle X\right\rangle _{\eta} & = & \sqrt{\frac{2\hbar}{m\omega_{x}}}\textrm{Re}(\eta),\\
\left\langle P_{x}\right\rangle _{\eta} & = & \sqrt{2m\hbar\omega_{x}}\textrm{Im}(\eta),\\
\Delta X_{\eta} & = & \sqrt{\frac{\hbar}{2m\omega_{x}}},
\end{eqnarray}
and where the global phase $\theta_{\eta}$ plays no physical role.
By Fourier transforming $\varphi(x,t=0)$, a parallel form for the
counterpart function in the momentum representation is obtained. The
momentum distribution of a coherent state simulates the initial preparation
given by Eq. (\ref{eq:MomDistr}). In fact, for the parallelism to
be complete, the width parameter $\sigma_{w}$ of the practical realization
must match that of a minimum-uncertainty wave packet. However, we
will see that our conclusions are robust when the width of the Gaussian
is varied. As it is relevant also to the analysis of many-body effects,
this issue will be dealt with in the next section. By now, we assume
that the preparation can be modeled by a coherent state and that $\eta$
is known from the experimental conditions. 

Switching to the new representation $\left|\tilde{\varphi}\right\rangle $,
we have $\left|\tilde{\varphi}(t=0)\right\rangle =D^{\dagger}(\beta)\left|\varphi(t=0)\right\rangle =D(-\beta)\left|\varphi(t=0)\right\rangle $.
Moreover, taking into account that the initial state can be expressed
as a displaced vacuum \cite{key-Louisell}, i.e., $\left|\varphi(t=0)\right\rangle =D(\eta)\left|0\right\rangle $,
we can write 

\begin{equation}
\left|\tilde{\varphi}(t=0)\right\rangle =D(-\beta)D(\eta)\left|0\right\rangle =e^{i\textrm{Im}(-\beta\eta^{\star})}D(\eta-\beta)\left|0\right\rangle =e^{i\textrm{Im}(-\beta\eta^{\star})}\left|\eta-\beta\right\rangle .
\end{equation}
Hence, the initial state corresponds to the coherent state $\left|\eta-\beta\right\rangle $
of the transformed Hamiltonian. Its time evolution is given by \cite{key-Louisell}

\begin{equation}
\left|\tilde{\varphi}(t)\right\rangle =e^{i\textrm{Im}(-\beta\eta^{\star})}\left|(\eta-\beta)e^{-i\omega_{x}t}\right\rangle .
\end{equation}
In particular, at the crossing, ($t=t_{c}$), we have for the mean
values of the coordinate and (displaced) momentum 

\begin{eqnarray}
q_{0} & = & \sqrt{\frac{2\hbar}{m\omega_{x}}}\textrm{Re}\left[(\eta-\beta)e^{-i\omega_{x}t_{c}}\right],\\
\hbar k_{x,c}+\frac{\alpha_{0}}{\hbar}m & = & \sqrt{2m\hbar\omega_{x}}\textrm{Im}\left[(\eta-\beta)e^{-i\omega_{x}t_{c}}\right]
\end{eqnarray}
where we have made use of the applicability of Eqs. (34) and (35)
to any coherent state. Combining the above equations, the optimum
displacement to introduce in the unitary transformation in Eq. (\ref{eq:eikonalUnitTransf})
is evaluated as 

\begin{eqnarray}
q_{0} & = & \sqrt{\frac{2}{m\omega_{x}^{2}}}\left[\hbar\omega_{x}\left|\eta\right|^{2}+\frac{\alpha_{0}}{\hbar}\sqrt{2m\hbar\omega_{x}}\textrm{Im}(\eta)-\left(\frac{(\hbar k_{x,c})^{2}}{2m}+\alpha_{0}k_{x,c}\right)\right]^{1/2}
\end{eqnarray}
Moreover, taking into account that 

\begin{equation}
\left\langle \eta\right|\frac{P_{x}^{2}}{2m}+\frac{\alpha_{0}P_{x}}{\hbar}+\frac{1}{2}m\omega_{x}^{2}X^{2}\left|\eta\right\rangle =\hbar\omega_{x}\left|\eta\right|^{2}+\frac{\alpha_{0}}{\hbar}\sqrt{2m\hbar\omega_{x}}\textrm{Im}(\eta),
\end{equation}
we recover the value of $q_{0}$ given by Eq. (\ref{eq:q0value}).
The complete agreement between the two presented derivations of $q_{0}$
is rooted in the quasi-classical character of the evolution of a coherent
state and in the crucial role played by the trajectory of the center
of the Gaussian wave packet in the effective model. That correspondence
is a proof of the validity of our simple approach. Again, the analogy
of our procedure with the eikonal approximation is apparent: the application
of the unitary transformation given by Eq. (\ref{eq:eikonalUnitTransf})
is equivalent to extract the phase of the coherent-state wave function
in the momentum representation. Namely, the term $\exp\left[-\frac{i}{\hbar}\left\langle X\right\rangle _{\eta}p\right]$
is taken out as a factor, and, the resulting reduced Hamiltonian is
cast into the LZ scenario.

It is of interest to discuss at this point the possibility of reaching
the adiabatic regime, i.e., of significantly diminishing $P_{LZ}^{(a)}=e^{-2\pi\left|\zeta\right|^{2}/\left|v\right|}$.
A reduction in $v$ can only be obtained via the decrease of $q_{0}$,
see Eq. (25). However, as previously stated, a significant value of
$q_{0}$ is needed for the consistency of our approach. Then, reducing
$v$ is not allowed, and, in order to diminish the probability of
transition between adiabatic states, one can only think of increasing
the offset, as found in the case of gravitational acceleration. However,
the implementation of this requirement, extracted from an approximate
description of the system which involves the translation to the crossing,
can introduce inconsistencies in our framework. Actually, a large
offset can induce transitions previous to the crossing, and, consequently,
invalidate the implemented approximation. Anyway, in that case, i.e.,
for a sufficiently large value of $\Omega$, a useful alternative
strategy to tackle the dynamics is the application of an adiabatic
approximation in the original Hamiltonian, with no simplification.
That procedure is more complex than the use of the translation technique,
but it is still feasible.

\section{Many-body effects}

The description of many-body interaction effects is required by the
considered practical conditions, which corresponded to the preparation
as a condensate. In a preliminary analysis of the experiments \cite{key-lzSOC},
the role of interaction effects in the appearance of the observed
features was evaluated. Given that, in the implemented setup, the
magnitude of the atomic-interaction energy was much smaller than the
kinetic energy, a single-particle approach was considered to be sufficiently
accurate to describe the dynamics. The agreement between the predictions
of the (single-particle) LZ model and the observed features validates
that conclusion. Here, we intend to go beyond the conditions corresponding
to the particular experimental realization and assess the robustness
of the single-particle approach as the magnitude of the interaction
is increased. The clarification of this aspect of the dynamics is
crucial for evaluating the potential applicability of the studied
transitions in different contexts, and, in particular, for assessing
their usefulness in strategies of control. In our analysis, we proceed
by reviewing first some general characteristics of the interaction.
Then, we present a quantitative evaluation of the corrections that
many-body effects introduce in the single-particle approach. In this
way, we delimit the range of safe applicability of the basic LZ scenario.

\subsection{General characterization of the interaction effects}

From the knowledge of the system dynamics provided by the present
study and by previous related work, the following general arguments
can be outlined: 

Due to the SOC, the system presents specific atomic-interaction characteristics.
In particular, there are different interaction strengths associated
with the spin combinations. They will be denoted as $g_{\eta,\eta^{\prime}}$,
where $\eta,\eta^{\prime}=1,2$ refer to the spins. As shown in previous
studies, depending on the SOC parameters and on the values of $g_{\eta,\eta^{\prime}}$,
the form of the wave-function of the ground state can display a varied
topology \cite{key-StringariSOCreview}. In the case of uniform confinement,
three phases, known as stripe, plane-wave (separate dressed-state),
and single-minimum phases, can appear \cite{key-StringariSOCreview,key-ZhangPhase}.
(See also \cite{key-finiteTempPhaseDiagram} for experimental studies
of finite-temperature phase diagrams). These phases persist when harmonic
confinement is considered. Apart from altering the form of the ground
state, many-body effects can modify the evolution of the initially
prepared state. Both aspects are relevant to the considered setup.
Their detailed description is quite complex in a general regime. Still,
the identification of some of their general characteristics can be
sufficient to evaluate their role in the present context. As in the
preliminary analysis of the experiments, we assume here that the realized
initial state is satisfactorily simulated by the product of one of
the spin states and an external state with a Gaussian distribution
of momenta. We aim at identifying differential aspects in the evolution
of this state associated with the inclusion of many-body effects. 

Important for our discussion is to recall some results of former work
on the application of a variational method to characterize the evolution
of a Bose-Einstein condensate. In the absence of spin-orbit coupling,
the use of a Gaussian ansatz with a set of dynamical variational parameters,
referring to the center and to the widths of the Gaussian, showed
that the evolution of the center is uncoupled from the dynamics of
the rest of the parameters \cite{key-CirZollGauss}. Moreover, the
atomic interaction was found to play no role in the dynamics of the
center, which presented no differences with that given by a single-particle
harmonic Hamiltonian. Many-body effects alter the dynamics of the
other parameters, i.e., the form of the Gaussian, and, consequently,
the distribution of momenta. These results, explicitly obtained for
a harmonic confinement, can be shown to be valid also for a linear
potential, and, therefore, for the case of gravitational acceleration. 

In recent studies \cite{key-collecmodeCenter}, the same variational
technique has been applied to analyze the evolution of spin-orbit
coupled Bose-Einstein condensates in the regime of separate dressed
states. The conclusions on the lack of coupling between the evolution
of the center of the Gaussian and the dynamics of other modes still
apply provided that the different interaction strengths relevant to
the SOC system do not significantly differ, specifically, when $g_{2}\equiv\left|(g_{12}-g_{11})/4\right|\ll g_{1}\equiv(g_{12}+g_{11})/4$,
($g_{11}\simeq g_{22}$). This restriction is fulfilled for the systems
analyzed in standard experimental realizations. As $g_{2}$ increases,
the evolution of the center becomes coupled with the dynamics of other
modes; moreover, nonlinear effects set in.

\subsection{Evaluation of the corrections to the single-particle description}

We turn now to set up a framework where a quantitative analysis of
many-body effects can be carried out. The starting point is the time-dependent
Gross-Pitaevskii equation

\begin{equation}
i\hbar\frac{\partial\mathbf{\boldsymbol{\Psi}_{s}}(\mathbf{r},t)}{\partial t}=\left[H+Ng_{int}\rho(\mathbf{r},t)\right]\mathbf{\boldsymbol{\Psi}_{s}}(\mathbf{r},t)
\end{equation}
where $H=-\frac{\hbar^{2}}{2m}\nabla_{\mathbf{r}}^{2}+V_{ex}(\mathbf{r})+\frac{\hbar\Omega}{2}\sigma_{x}+\left(\frac{\hbar\delta_{0}}{2}-i\alpha_{0}\frac{\partial}{\partial x}\right)\sigma_{z}$
with $V_{ex}(\mathbf{r})=\frac{1}{2}m(\omega_{x}^{2}x^{2}+\omega_{y}^{2}y^{2}+\omega_{z}^{2}z^{2})$,
and $\mathbf{\boldsymbol{\Psi}_{s}}\equiv\begin{pmatrix}\Psi_{+}(\mathbf{r},t)\\
\Psi_{-}(\mathbf{r},t)
\end{pmatrix}$ denotes a two-component spinor wave function in the coordinate representation,
$N$ is the number of particles, and $\rho(\mathbf{r},t)=\left|\Psi_{+}\right|^{2}+\left|\Psi_{-}\right|^{2}$.
Note that a three-dimensional description is necessary. For simplicity,
we consider that the involved interaction strengths are equal, specifically,
$g_{int}\equiv g_{11}=g_{22}=g_{12}$. The implications of dealing
with different scattering lengths will be discussed. In our procedure
to solve Eq. (43), we apply first a variational method in the line
of previous studies on SOC condensates. Despite the restrictions associated
with the lack of generality of the proposed ansatz, the results will
provide us with useful clues to setting up a scheme of general validity.
From that approach, the predictions of the single-particle study will
be recovered.

\subsubsection{The variational approach}

Our method combines elements of previous applications of variational
techniques to related systems. First, we propose an ansatz for the
evolution of the center of the packet via the sequence of unitary
transformations 

\begin{equation}
\mathbf{\mathbf{\boldsymbol{\Psi}_{s}}}(\mathbf{r},t)=U_{4}(t)U_{3}(t)\boldsymbol{\Psi}(\mathbf{r},t),
\end{equation}
where 

\begin{equation}
U_{3}(t)=\exp\left[-\frac{i}{\hbar}\mathbf{r}_{c}(t)\mathbf{P}\right],
\end{equation}

\begin{equation}
U_{4}(t)=\exp\left[-\frac{i}{\hbar}\mathbf{p}_{c}(t)\mathbf{R}\right],
\end{equation}
with $\mathbf{P}$ and $\mathbf{R}$ being the particle position and
momentum vector operators. For the transformed wave function, we propose
the ansatz
\begin{equation}
\boldsymbol{\Psi}(\mathbf{r},t)=\Phi_{0}(\mathbf{r},t)\begin{pmatrix}c_{+}(t)\\
c_{-}(t)
\end{pmatrix}.
\end{equation}
Note that, as no spatial dependence is allowed for the coefficients
$c_{+}$ and $c_{-}$, the proposal lacks generality. Still, as opposed
to former applications of the variational techniques \cite{key-collecmodeCenter},
no restrictions are imposed on the time dependence of the spin \emph{coefficients}
and on the functional form of $\Phi_{0}(\mathbf{r},t)$. We shall
now determine the functions $\mathbf{r}_{c}(t)$, $\mathbf{p}_{c}(t)$,
$\Phi_{0}(\mathbf{r},t),$ $c_{+}(t)$ and $c_{-}(t)$ from the stationary
action principle 
\[
\delta\int_{t_{0}}^{t}dt^{\prime}\int d^{3}\mathbf{r\,\mathbf{\mathbf{\boldsymbol{\Psi}_{s}^{\dagger}}}}[H-i\hbar\frac{\partial}{\partial t^{\prime}}+\frac{1}{2}Ng_{int}\rho(\mathbf{r},t^{\prime})]\mathbf{\mathbf{\boldsymbol{\Psi}_{s}}}=0,
\]
where $\mathbf{\mathbf{\boldsymbol{\Psi}_{s}^{\dagger}}}\equiv(\Psi_{+}^{\star},\Psi_{-}^{\star}$)
is the transpose complex conjugate spinor wave function. The solution
must satisfy the supposedly initial known form $\mathbf{\mathbf{\boldsymbol{\Psi}_{s}}}(\mathbf{r},t_{0})$
of this spinor wavefunction. If $\mathbf{r}_{c}(t_{0})$ and $\mathbf{p}_{c}(t_{0})$
are chosen such that 
\begin{eqnarray}
\left\langle \Phi_{0}(t_{0})\right|\mathbf{R}\left|\Phi_{0}(t_{0})\right\rangle  & = & \mathbf{0}\\
\left\langle \Phi_{0}(t_{0})\right|\mathbf{P}\left|\Phi_{0}(t_{0})\right\rangle  & = & \mathbf{0}.
\end{eqnarray}
the variational solutions for $\mathbf{r}_{c}(t)$ and $\mathbf{p}_{c}(t)$,
intended to define the trajectory of the packet center, are given
by the set of equations

\begin{eqnarray}
\begin{pmatrix}\dot{x}_{c}\\
\dot{y}_{c}\\
\dot{z}_{c}
\end{pmatrix} & = & \frac{1}{m}\begin{pmatrix}p_{c,x}\\
p_{c,y}\\
p_{c,z}
\end{pmatrix}+\left[\left|c_{+}\right|^{2}-\left|c_{-}\right|^{2}\right]\frac{\alpha_{0}}{\hbar}\begin{pmatrix}1\\
0\\
0
\end{pmatrix},\\
\begin{pmatrix}\dot{p}_{c,x}\\
\dot{p}_{c,y}\\
\dot{p}_{c,z}
\end{pmatrix} & = & -m\begin{pmatrix}\omega_{x}^{2}x_{c}\\
\omega_{y}^{2}y_{c}\\
\omega_{z}^{2}z_{c}
\end{pmatrix},
\end{eqnarray}
 The variational solution for $\Phi_{0}(\mathbf{r},t)$ is provided
by the equation 

\begin{equation}
i\hbar\frac{\partial\Phi_{0}}{\partial t}=\left[E_{c}-\dot{\mathbf{r}}_{c}\mathbf{p}_{c}-\frac{\hbar^{2}}{2m}\nabla_{\mathbf{r}}^{2}+V_{ex}+Ng_{int}\left|\Phi_{0}\right|^{2}\right]\Phi_{0},
\end{equation}
with $E_{c}(t)=\frac{1}{2m}\mathbf{p}_{c}^{2}+\frac{1}{2}m\left(\omega_{x}^{2}x_{c}^{2}+\omega_{y}^{2}y_{c}^{2}+\omega_{z}^{2}z_{c}^{2}\right)+\left(\left|c_{+}\right|^{2}-\left|c_{-}\right|^{2}\right)\frac{\alpha_{0}}{\hbar}p_{c,x}$
. Finally, the evolution of the spin coefficients $c_{+}(t)$ and
$c_{-}(t)$ is found to be given by

\begin{eqnarray}
i\hbar\frac{\partial}{\partial t}\begin{pmatrix}c_{+}\\
c_{-}
\end{pmatrix} & = & \left[\frac{\hbar\Omega}{2}\sigma_{x}+\Bigl(\frac{\hbar\delta_{0}}{2}+\frac{\alpha_{0}}{\hbar}p_{c,x}\Bigr)\sigma_{z}\right]\begin{pmatrix}c_{+}\\
c_{-}
\end{pmatrix}.
\end{eqnarray}
Hence, the application of the variational method with an ansatz with
spatial-independent coefficients $c_{+}$ and $c_{-}$ leads exactly
to the LZ scenario. It is worth emphasizing that the functions $\mathbf{r}_{c}(t)$
and $\mathbf{p}_{c}(t)$ {[}see Eqs. (50) and (51){]}, describe the
dynamics of a classical harmonic oscillator modified by a spin-dependent
driving term. Additionally, the dynamics of $\Phi_{0}$ parallels
that of the wave function of a condensate in the absence of SOC in
a harmonic trap. Indeed, Eq. (52) can be cast into the standard form
of the time-dependent Gross-Pitaevskii equation by incorporating the
extra (time-dependent, coordinate-independent terms) $E_{c}$ and
$\dot{\mathbf{r}}_{c}\mathbf{p}_{c}$ as a phase $\Phi_{0}(\mathbf{r},t)\rightarrow\Phi_{0}(\mathbf{r},t)e^{-\frac{i}{\hbar}\int_{t_{0}}^{t}(E_{c}-\dot{\mathbf{r}}_{c}\mathbf{p}_{c})dt^{\prime}}$.
In order to simplify, the discussion of our results, the formal elimination
of those terms will be considered in the following. Hence, without
changing the notation for the transformed wave function, we rewrite
Eq. (52) as 

\begin{equation}
i\hbar\frac{\partial\Phi_{0}}{\partial t}=\left[-\frac{\hbar^{2}}{2m}\nabla_{\mathbf{r}}^{2}+V_{ex}+Ng_{int}\left|\Phi_{0}\right|^{2}\right]\Phi_{0},
\end{equation}

\subsubsection{Generalization of the method: results for an ansatz with general
validity }

Now, taking as starting point the scheme provided by the variational
method, we turn to set up a framework of general validity. Again,
we use the sequence of unitary transformations given by Eqs. (44),
(45), (46), but, now, a generalized proposal is made for the trajectory
of the packet center: although Eq. (51) still applies, Eq. (50) is
modified as 

\begin{eqnarray}
\begin{pmatrix}\dot{x}_{c}\\
\dot{y}_{c}\\
\dot{z}_{c}
\end{pmatrix} & = & \frac{1}{m}\begin{pmatrix}p_{c,x}\\
p_{c,y}\\
p_{c,z}
\end{pmatrix}+\left[\left\langle \left|c_{+}\right|^{2}\right\rangle -\left\langle \left|c_{-}\right|^{2}\right\rangle \right]\frac{\alpha_{0}}{\hbar}\begin{pmatrix}1\\
0\\
0
\end{pmatrix},
\end{eqnarray}
 with $c_{+}(\mathbf{r},t)$ and $c_{-}(\mathbf{r},t)$ being functions,
with spatial dependence, that we use in a different (general) characterization
of the spinor wave function: without loss of generality, we write
\begin{equation}
\mathbf{\boldsymbol{\Psi}}(\mathbf{r},t)=\Phi_{0}(\mathbf{r},t)\begin{pmatrix}c_{+}(\mathbf{r},t)\\
c_{-}(\mathbf{r},t)
\end{pmatrix}.
\end{equation}
The function $\Phi_{0}(\mathbf{r},t)$ is now imposed to obey Eq.
(52), which does not imply a restriction since $c_{+}$ and $c_{-}$
incorporate spatial dependence. Note that $\left\langle \left|c_{\pm}\right|^{2}\right\rangle \equiv\left\langle \Phi_{0}\right|\left|c_{\pm}\right|^{2}\left|\Phi_{0}\right\rangle $.
Then, through the application of the sequence of unitary transformations
and the use of our ansatz for the transformed wave function in Eq.
(43), the evolution of the spin coefficients is found to be given
by 

\begin{eqnarray}
i\hbar\frac{\partial}{\partial t}\begin{pmatrix}c_{+}\\
c_{-}
\end{pmatrix} & = & \Biggl\{-\frac{\hbar^{2}}{2m}\nabla_{\mathbf{r}}^{2}-\frac{\hbar^{2}}{m\Phi_{0}}(\boldsymbol{\nabla}_{\mathbf{r}}\Phi_{0})\boldsymbol{\nabla}_{\mathbf{r}}+Ng_{int}\left|\Phi_{0}\right|^{2}\Bigl(\left|c_{+}\right|^{2}+\left|c_{-}\right|^{2}-1\Bigr)+\nonumber \\
 &  & -\frac{i\alpha_{0}}{\Phi_{0}}\frac{\partial\Phi_{0}}{\partial x}\Bigl[\sigma_{z}-\Bigl(\Bigl\langle\left|c_{+}\right|^{2}\Bigr\rangle-\Bigl\langle\left|c_{-}\right|^{2}\Bigr\rangle\Bigr)\Bigr]-i\alpha_{0}\sigma_{z}\frac{\partial}{\partial x}+\nonumber \\
 &  & \frac{\hbar\Omega}{2}\sigma_{x}+\Bigl(\frac{\hbar\delta_{0}}{2}+\frac{\alpha_{0}}{\hbar}p_{c,x}\Bigr)\sigma_{z}\Biggr\}\begin{pmatrix}c_{+}\\
c_{-}
\end{pmatrix}.
\end{eqnarray}
Note that, in setting up our framework, no approximations have been
made apart from the mean-field approach that leads to the Gross-Pitaevskii
equation. Despite the complexity of the description, the following
basic characteristics of the role of many-body effects in the emergence
of the LZ scenario can be identified:

i) Interaction effects enter directly Eq. (54): they affect the form
of $\Phi_{0}$. In turn, as $\Phi_{0}$ varies, the dynamics of the
spin coefficients, given by Eq. (57), is modified. The indirect role
of many-body effects in the trajectory of the packet center is also
apparent. Moreover, since Eq. (54) has the same functional structure
as that corresponding to the dynamics of a condensate in the absence
of SOC, a parallelism can be traced between both systems. This implies
the potential utility in our system of the well-known characterization
of the condensate without SOC. 

ii) The departure of Eq. (57) from the standard LZ scenario is rooted
in the terms that incorporate spatial derivatives, which are actually
relevant when the functions $c_{+}$ and $c_{-}$ are nonuniform,
and in the term that includes $\left|\Phi_{0}\right|^{2}$. As the
preparation of the system in the experiments can be approximated by
the product of one of the spin states and the fundamental state of
a condensate in a harmonic trap, $c_{+}$ and $c_{-}$ are initially
uniform. The spatial derivatives are activated only when, due to the
coupling term $\frac{\hbar\Omega}{2}\sigma_{x}$, non-uniformity sets
in. Specifically, the role of non-uniformity initiates when the transfer
of population modifies the initial value of the difference $\Bigl\langle\left|c_{+}\right|^{2}\Bigr\rangle-\Bigl\langle\left|c_{-}\right|^{2}\Bigr\rangle$,
which is $1$ or $-1$, depending on the prepared spin state. The
same argument applies to the term with $\left|\Phi_{0}\right|^{2}$.
Consequently, we conclude that  it is only near the crossing that
the form of $\Phi_{0}$, in particular, its evolution, due to interaction
effects, becomes relevant to the dynamics of the spin functions. 

iii) Let us see that the form of the expressions obtained for the
probability of transition in the single-particle approach is robust
against many-body effects. Important to this point is to recall that
the nonuniform terms that affect the evolution of the spin functions
are also present in the single-particle description, which can be
simply recovered by taking $g_{int}=0$ in Eq. (54) and Eq. (57).
Spatial variation of the wave function, given then by the Schrödinger
equation, enters the corrections to the basic LZ model. Even when
the (position-dependent) fundamental state of the harmonic trap is
prepared and no changes take place in $\Phi_{0}$, spatial derivatives
are activated as soon as the transfer of population becomes relevant.
As previously discussed, those corrections are small provided that
the coupling becomes effective only very near the crossing, which
is in turn guaranteed by the asymptotic character of the LZ scenario.
The order of magnitude of the corrections does not significantly varies
when interaction is considered, and, therefore, the form of the LZ
formula is robust. It is clear that these conclusions give validity
to our use, in the previous section, of a coherent state in the simulation
of the single-particle dynamics irrespective of the actual value of
the Gaussian width.

Our framework can be easily adapted to deal with the case of gravitational
acceleration. Indeed, it is shown that the result referring to the
exact validity of the LZ model for gravitational acceleration in the
single-particle case can be recovered. Switching to the many-body
scenario, one can conjecture that, due to the lack of confinement
in the gravitational acceleration, the (repulsive) interaction effects
must be less important that in the case of harmonic trapping: the
unimpeded spreading of the wave function implies smaller spatial derivatives.

iv) The asymptotic character of the LZ predictions must be stressed:
the LZ formula connects the initial and final configurations, which
are both very far from the crossing. The compact form of the transition
probability is actually due to these characteristics. It is again
apparent that the magnitude of the coupling term $\frac{\hbar\Omega}{2}\sigma_{x}$
determines the definition of the asymptotic regions. As $\Omega$
increases, larger distances from the crossing are required for the
initial and final configurations. This condition implies limitations
on the practical realization. Here, it is worth mentioning the results
of \cite{key-nonAdiabaManyBoSOC}, where departures from the LZ predictions
were observed for increasing values of $\Omega$; also in agreement
with our results, deformation of the wave packet due to interaction
effects were observed.

v) Our approach allows us to evaluate the effect of interactions on
the evolution of the center of the wave packet. This point is crucial
since it is the dynamics of the mean value of the momentum, i.e.,
the trajectory of the center of the packet, that determines the sweeping-rate
parameter of the simplified description of the transitions. From Eqs.
(51) and (55), one can conclude that, again, it is near the crossing,
where interaction effects become effective, that the spin-induced
driving of the trajectory can be significant. A modification of the
LZ velocity can be expected, but, still, the applicability of the
LZ formula for the probability of transition is guaranteed provided
that the asymptotic conditions are fulfilled. The analysis of Eq.
(55) shows that the relative importance of the spin-dependent driving
of the center-of-the-packet variables decreases as the quotient between
$\frac{\alpha_{0}}{\hbar}$ and the magnitude of $\frac{p_{c,x}}{m}$
diminishes. The experimental conditions \cite{key-lzSOC} indeed correspond
to a very small value of that quotient. Indeed, a perturbative treatment
of the terms additional to the primary LZ scenario shows that it is
only beyond first order that corrections to the spin population given
by the basic LZ model appear. A detailed presentation of the technical
details of that treatment will be given in future work.

From the above general picture of the dynamics, one can infer the
sound applicability of the basic LZ model to the many-particle scenario
in standard conditions. Departures from the above description appear
when the magnitudes of the different scattering lengths (diagonal
and non-diagonal) relevant to the problem are significantly different.
It can be shown that then the spin functions $c_{+}$ and $c_{-}$
enter the equation for the dynamics of $\Phi_{0}(\mathbf{r},t)$,
i.e., Eq. (54) is significantly modified. Moreover, the different
scattering lengths appear explicitly in the equation that gives the
evolution of the spin functions. We intend to evaluate in future work
if the basic structure of the LZ model can still be traced in this
more complex scenario.

Finally, it is pertinent to stress the substantial differences between
the system analyzed here and other variations of the LZ scenario,
relevant to different contexts, which present a significant modification
of the single-particle output. For instance, it is worth mentioning
studies which have focused on the effect of a nonlinear coupling \cite{key-Niu}
and of a nonlinear sweeping rate \cite{key-PhysLettA}. There, the
term \emph{nonlinear} refers to dependence on the spin populations.
As opposed to those models, in our system,  no dependence of the coupling
term (or of the sweeping rate) on the spin populations is present.
Again, we stress that departures from this picture can be expected
when different (spin-dependent) scattering lengths are considered.

\section{Concluding remarks}

In our study, the LZ model has been generalized along two lines. First,
we have enlarged the dimensionality of the system by including the
external variables. This has been done for two forms of the external
forces: linear and quadratic potentials have been considered. In both
cases, the emergence of the basic LZ model has been analytically traced.
Second, we have incorporated many-body effects. Given the characteristics
of the atomic-interaction strengths, the many-body dynamics does not
alter the applicability of the model: the single-particle description
has been found to be robust when atomic-interaction effects are incorporated.

Our analysis has uncovered the crucial role played by the properties
of the initial preparation of the system in the applicability of the
LZ description. We have found that, for the LZ model to be valid,
the momentum components of the initial state must correspond to large
values of that observable. The measurement on the system must also
be carried out when the momentum has reached again a large-magnitude
value, i.e., the effective LZ ramp must be ended sufficiently far
from the crossing. These requirements were fulfilled in the experimental
realization. 

The above arguments are also relevant to the observed robust character
of the applicability of the model when the characteristics of the
acceleration methods are varied. The use of different acceleration
schemes only alters the dynamics outside the crossing, which is relevant
merely to determine the rate of variation of the momentum at the transition.
As a side effect of the validity of the LZ approach, the lack of dependence
of the transition probability on the detuning has been explained.
Limitations on reaching an adiabatic regime for the transitions have
been revealed: only by increasing the offset parameter, the transition
between adiabatic states can be inhibited in practice. 

The exact analytical results obtained for the scheme based on gravitational
acceleration can find additional applicability in the precise characterization
of the evolution of different initial states outside the asymptotic
regime. The use of the LZ approach is frequently limited to the evaluation
of asymptotic probabilities of transition. From our analytical approach,
a description that goes beyond that framework is feasible. 

Apart from a more complete explanation of the experimental results,
our analytical results can provide clues to steer the dynamics. The
study gives additional support to the strategy of control proposed
in Ref. \cite{key-lzSOC}: the spin polarization of the output can
be tuned using the offset and the rate of splitting variation as parameters
of control. It is also apparent that the study can be relevant to
strategies to manipulate the system based on the variation of the
SOC parameters \cite{key-SpielmanControl,key-ZhangSciRep,key-Llorente}.
Indeed, it is pertinent to take into account the potential role of
the trapping potential as a mechanism of acceleration whenever a non-equilibrium
situation is tackled. The analysis  can provide clues to advancing
in the characterization of aspects of the dynamics of spin-orbit coupled
Bose-Einstein condensates like the coupling of collective modes \cite{key-collecmodeCenter,key-StringariSOC1,key-StringariSOC2,key-StringariSOCreview,key-zhangDipole}
and the differential effects of the diverse ground-state characteristics.

\section*{Acknowledgments}

One of us (JMGL) acknowledges the support from Ministerio de Economía
y Competitividad (Grant No. FIS2013-41532-P) and from the European
Regional Development Fund.

\end{document}